# Pressure dependence of structural, elastic, electronic, thermodynamic, and optical properties of van der Waals-type NaSn$_2$P$_2$ pnictide superconductor: insights from DFT study

F. Parvin and S.H. Naqib*

Department of Physics, University of Rajshahi, Rajshahi-6205, Bangladesh

*Corresponding author salehnaqib@yahoo.com

**Abstract**

NaSn$_2$P$_2$ is a recently discovered superconducting system belonging to a particular class of materials with van der Waals (vdW) structure. There is enormous interest in vdW compounds because of their intriguing electrical, optical, chemical, thermal, and superconducting state properties. We have studied the pressure dependent structural, thermo-physical, electronic band structure, and superconducting state properties of this quasi-two dimensional system in details for the first time via ab initio technique. The optical properties are also investigated for different electric field polarizations for the first time. Structural, electronic, and optical properties were explored via density functional theory (DFT) calculations. Thermal properties were investigated using the quasi-harmonic Debye model. NaSn$_2$P$_2$ is found to be mechanically stable in the pressure range 0 – 3.0 GPa. The elastic anisotropy indices point towards high level of mechanical and bonding anisotropy in NaSn$_2$P$_2$ consistent with its highly layered structure. The elastic constants, moduli, and Debye temperature ($\theta_D$) show non-monotonic variation with pressure, particularly close to 1.0 GPa. The pressure dependent superconducting transition temperature, $T_c$, of NaSn$_2$P$_2$ is predicted to vary strongly with the pressure dependent variation of $\theta_D$. The electronic energy dispersion curves, $E(k)$, reveal high level of direction dependence; the effective mass of charge carries are particularly high for the out-of-plane ($c$-axis) charge transport. The optical parameters compliment the underlying electronic energy density of states features and are weakly dependent on the polarization of the incident electric field. The reflectivity of NaSn$_2$P$_2$ is very high in the visible region and remains quite high and non-selective over an extended energy range in the ultraviolet region. The absorption coefficient is also high in the mid-ultraviolet band. All these optical features render NaSn$_2$P$_2$ suitable for optoelectronic device applications.

# 1. Introduction

Interests in lower dimensional electronic systems including two-dimensional (2D) compounds and van der Waals (vdW) solids are growing rapidly across various strata of scientific and engineering disciplines due to their remarkable thermoelectric, electronic, optical, thermal, and chemical properties [1]. In recent years micromechanical exfoliation techniques have been adopted widely for rapid material characterization and demonstration of novel device ideas based on these layered 2D systems [1]. Furthermore, significant advances have been attained in large-scale homogeneous and heterogeneous synthesis of these materials [1, 2]. Recent discovery of superconductivity in vdW solids is another significant landmark [3]. Superconductivity in low dimensional systems is of fundamental interest because of their exotic electronic structure, possibility of easy atomic intercalation and novel electronic and magnetic orders [4 – 7]. Low dimensionality often facilitates strong Cooper pairing and can gives rise to unconventional superconducting order parameter [7 – 9].

Very recently, superconductivity with a transition temperature in the range 1.2 K to 1.6 K was found in the SnAs based layered compound $NaSn_2As_2$ belonging to a new class of vdW-type superconductor [3 – 5]. At the same time, superconductivity was also discovered in Na deficient isostructural vdW type compound $Na_{1-x}Sn_2P_2$ with a $T_c$ of 2.0 K [4]. It was also demonstrated that Na-doping at the Sn atomic site in $NaSn_2As_2$ ($Na_{1+x}Sn_{2-x}As_2$) is effective in enhancing superconductivity, leading to a $T_c$ = 2.1 K for $x$= 0.4. The difference in $T_c$ for $NaSn_2As_2$, $Na_{1-x}Sn_2P_2$, and $Na_{1+x}Sn_{2-x}As_2$ has been attributed mainly to the difference in the electronic density of states (EDOS) [3 – 5] at the Fermi level, $N(E_F)$.

$NaSn_2As_2$ was found to crystallize in a trigonal $R\bar{3}m$ unit cell, consisting of two layers of a buckled honeycomb network of SnAs unit bound by the vdW forces and separated by the Na ions [3, 10]. Due to the presence of vdW gap between the SnAs layers, this compound can be readily exfoliated through both mechanical and liquid-phase methods [10, 11]. It is interesting to note that there are various other structural analogues with conducting tin-pnictide (SnPn) layers, including $Sn_4Pn_3$ [12, 13] and ASnPn [14 – 18], as well as $ASn_2Pn_2$ [19 – 21], where A symbolizes alkali or alkaline earth metal atoms. $Sn_4Pn_3$ exhibits superconductivity with $T_c$ = 1.2– 1.3 K [22, 23]. Moreover, ASnPn compounds possess a number of thermal and electronic properties highly pertinent for thermoelectric device applications, e.g., low lattice thermal conductivity and suitability for band structure manipulation [24, 25]. All these features signify high level of functionalities of these vdW type layered compounds.

Information regarding the elastic properties and mechanical anisotropy are required to unlock the full potential of a compound for possible applications. Furthermore, study of optical properties provides with useful information that can assist us in designing optoelectronic device appliances. The energy dependent optical parameters also supplement the results of electronic band structure calculations. Study of thermal properties reveals how a compound behaves at different temperatures and pressures. Application of pressure modifies the physical properties of compounds. Pressure dependent elastic and band structure calculations are important for the understanding of bonding character, electronic density of states and superconductivity in materials [26, 27]. The elastic constants, moduli, and compliances are intimately related to the bonding strengths of a crystal. The bonding and structural features determine the phonon spectrum, and therefore, set the energy scale for superconductivity. Moreover, study of elastic anisotropy indices provides us with valuable information regarding the possible crystal defects and their dynamics and viable mechanical failure modes of crystalline solids under different types of external stress. The electronic band structure and the electronic energy density of states at the Fermi energy are also important for emergence of superconductivity in a material.

As far as we are aware of, none of the physical properties mentioned in the preceding paragraph has been investigated in details for the van der Waals-type $NaSn_2P_2$ pnictide yet. We wish to address these issues in this paper. All the calculations performed in this study were done via the density functional theory (DFT) based *ab*-initio procedure.

The rest of this work has been organized as follows. Section 2 describes the methodology employed in this study. Results of our calculations and analyses are given and discussed in Section 3. Finally, important conclusions are summarized in Section 4.

## 2. Computational Methodology

Perhaps the most popular approach to first-principles calculations of structural and electronic properties of crystalline solids is the DFT with periodic Bloch boundary conditions. In this particular approach the ground state properties of the compound is found by solving the Kohn-famous Sham equation [28] which takes into account of both exchange and correlation terms in the total energy of the system. Selection of the proper scheme for exchange correlations is an important part for reliable implementation of the *ab*-initio calculations. We have checked various schemes for the exchange correlations (including GGA-PBE) to optimize the geometry of $NaSn_2P_2$ pnictide and have found that LDA (local density approximation) yields by far the best result as far as the crystal structure is concerned. Therefore, we have used the LDA to model electron exchange correlations as contained within the CAmbridge Serial Total Energy Package (CASTEP) [29] designed particularly to implement quantum mechanical DFT based calculations. Ultra-soft Vanderbilt-type pseudopotentials were used to calculate the electron-ion interactions. This scheme relaxes the norm-conserving constraint but produces a smooth and computation friendly pseudopotential which saves computational time significantly without affecting the accuracy appreciably [30]. Density mixing has been used to the electronic structure and the Broyden Fletcher Goldfarb Shanno (BFGS) geometry optimization [31] has been adopted to optimize the crystal structure for the given symmetry (Trigonal, space group, $R\bar{3}m$ (No. 166), for $NaSn_2P_2$, to be specific). The following electronic orbitals are considered for Na, Sn and P atoms to construct the valence and the conduction states, respectively: $Na[3s^1]$, $Sn[5s^25p^2]$ and $P[3s^23p^3]$. Periodic boundary conditions are employed to determine the total energies of each crystal cell. The trial wave functions are expanded in terms of plane wave basis set. The cut-off energy for the plane wave expansion has been taken as 750 eV. *k*-point sampling within the reciprocal space (Brillouin zone (BZ)) for the compound under investigation has been done with a dense mesh of 14×14×2 special points using the Monkhorst-Pack grid scheme [32]. These selections of cut-off energy and *k*-point set ensure high level of convergence of the total energy versus cell volume calculations. During geometry optimization, the following tolerance levels have been maintained; the self-consistent convergence thresholds of $10^{-6}$ eV atom$^{-1}$ for the total energy; 0.03 eV Å$^{-1}$ for the maximum force; 0.05GPa for maximum stress; and $10^{-3}$ Å for maximum lattice displacement. Elastic constants were estimated by the 'stress-strain' method as contained in the CASTEP program. The polycrystalline bulk modulus, *B* and the shear modulus, *G* were estimated from the calculated single crystal elastic constants expressed as $C_{ij}$. The electronic band structure features have been obtained from the theoretically optimized structure of NaSn$_2$P$_2$. Energy dependent optical constants spectra were obtained by considering the electronic transition probabilities among different electronic bands. The imaginary part, $\varepsilon_2(\omega)$, of the complex dielectric function has been estimated within the momentum representation of matrix elements between occupied and unoccupied electronic states in the valence and conduction states, respectively, by employing the CASTEP supported formula which can be expressed as

$$\varepsilon_2(\omega) = \frac{2e^2\pi}{\Omega\varepsilon_0} \sum_{k,v,c} \left|\left\langle \psi_k^c \left| \hat{u}\cdot\vec{r} \right| \psi_k^v \right\rangle\right|^2 \delta(E_k^c - E_k^v - E) \quad (1)$$

In this expression based on time dependent perturbation theory, $\Omega$ is the volume of the unit cell, $\omega$ is the angular frequency (or equivalently energy) of the incident electromagnetic wave (photon), *e* is the electronic charge, $\psi_k^c$ and $\psi_k^v$ are the conduction and valence band state functions with band index *k*, respectively. The delta function forces conservation of energy and momentum to be held during the optical transition. The Kramers-Kronig transformations connect the real part of the dielectric function $\varepsilon_1(\omega)$ to the imaginary part $\varepsilon_2(\omega)$ and can be readily used to obtain $\varepsilon_1(\omega)$ once $\varepsilon_2(\omega)$ is known. All the other optical parameters are related in one way or other to $\varepsilon_1(\omega)$ and $\varepsilon_2(\omega)$, and can be calculated from these. This procedure has been applied by a large number of earlier studies to reliably calculate all the energy/frequency dependent optical parameters [33 – 37].

To explore the pressure and temperature dependent thermodynamic, elastic, and structural properties, we have employed the energy-volume data calculated from the third-order Birch-Murnaghan equation of state [38] using the zero temperature and zero pressure equilibrium values of energy, volume, and bulk modulus obtained through the DFT calculations together with quasi-harmonic Debye model where appropriate.

## 3. Results and Analysis

*3.1. Structural and elastic properties*

The schematic crystal structure of van der Waals-type NaSn$_2$P$_2$ pnictide is displayed in Fig. 1. The optimized lattice parameters including the cell volume are given in Table 1. We have included results from published source [4] for comparison. The structural parameters have been calculated at different values of applied hydrostatic pressure up to 3.0 GPa. It is seen that the lattice parameter along *c*-axis is significantly larger compared to those along crystallographic *a*- and *b*-directions. This readily illustrates the highly layered character of NaSn$_2$P$_2$ structure. The geometrical optimization was performed by minimizing the energy of the crystal structure with respect to the volume of the unit cell. The structural ground state therefore, yields the fully relaxed lattice parameters of NaSn$_2$P$_2$. It should be noted that the experimental lattice parameters of NaSn$_2$P$_2$ was extracted from the X-ray diffraction profile at room temperature [4]. The theoretically estimated lattice parameters and cell volume, on the other hand corresponds to the ground state at absolute zero temperature. Therefore, the calculated values are naturally somewhat smaller than the experimental ones. The deviations between experimental and theoretical values become even more insignificant if the weak bondings among the atoms in NaSn$_2$P$_2$ (see the subsequent analyses) are considered which result in large values of linear and volume thermal expansion coefficients.

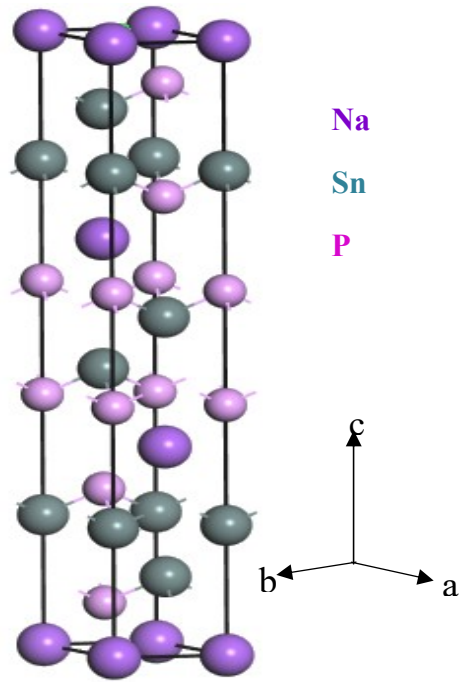

**Fig. 1.** Schematic crystal structure of NaSn$_2$P$_2$. The crystallographic directions are shown.

**Table 1**
Optimized lattice parameters and cell volumes of NaSn$_2$P$_2$ under different applied pressures.

| $P$ (GPa) | $a$ (Å) | $c$ (Å) | $V$ (Å$^3$) | Ref. |
|---|---|---|---|---|
| 0.0 | 3.8130 | 26.4673 | 333.25 | This |
| 0.0 | 3.8847 | 27.1766 | - | [4] |
| 0.5 | 3.8039 | 26.3318 | 329.97 | This |
| 1.0 | 3.7947 | 26.1890 | 326.60 | This |
| 1.5 | 3.7878 | 26.0264 | 323.38 | This |
| 2.0 | 3.7804 | 25.8891 | 320.43 | This |
| 2.5 | 3.7732 | 25.7769 | 317.82 | This |
| 3.0 | 3.7662 | 25.6499 | 315.08 | This |

NaSn$_2$P$_2$ assumes trigonal crystal structure. Therefore, from symmetry arguments, this layered vdW compound has six independent single crystal elastic/stiffness constants: $C_{11}$, $C_{12}$, $C_{13}$, $C_{14}$, $C_{33}$, and $C_{44}$. The stiffness constant $C_{66}$ depends on the two other independent stiffness constants $C_{11}$ and $C_{12}$ via the expression, $C_{66} = (C_{11} - C_{12})/2$. All the calculated elastic stiffness constants at different pressures are presented in Table 2. There is no prior report on these parameters for NaSn$_2$P$_2$, therefore, the reported values herein should serve as reference for future investigations.

**Table 2**
Single crystal elastic constants of NaSn$_2$P$_2$ (all in GPa).

| Pressure | $C_{11}$ | $C_{12}$ | $C_{13}$ | $C_{14}$ | $C_{33}$ | $C_{44}$ | $C_{66} = [(C_{11}-C_{12})/2]$ | Ref. |
|---|---|---|---|---|---|---|---|---|
| 0.0 | 115.59 | 43.48 | 19.51 | 3.10 | 73.40 | 18.65 | 36.06 | This |
| 0.5 | 119.43 | 49.21 | 23.92 | 4.53 | 70.22 | 17.67 | 35.11 | This |
| 1.0 | 117.78 | 57.98 | 26.93 | 12.68 | 77.12 | 8.83 | 29.90 | This |
| 1.5 | 127.44 | 50.89 | 29.91 | 0.30 | 84.21 | 25.31 | 38.28 | This |
| 2.0 | 129.85 | 52.04 | 30.21 | -0.11 | 86.95 | 26.59 | 38.91 | This |
| 2.5 | 133.29 | 55.00 | 34.07 | -0.62 | 87.27 | 25.17 | 39.15 | This |
| 3.0 | 135.19 | 56.17 | 34.84 | -0.69 | 89.85 | 28.19 | 39.51 | This |

It is seen from the above table that the calculated elastic tensors are relatively small and vary non-monotonically with increasing pressure. Among the six independent elastic constants, $C_{11}$ and $C_{33}$ determine the stiffness for axial compression. Crystal symmetry of NaSn$_2$P$_2$ implies that $C_{11} = C_{22}$. At all pressures $C_{11}$ is significantly higher than $C_{33}$. This shows that the crystal appears much stiffer against uniaxial stress along $a$- and $b$-axis in comparison to the stress applied along the $c$-axis. Such behavior is related to the layered character of the crystal structure and anisotropic nature of underlying chemical bonding. The other four elastic constants are related to the shear dominated responses. $C_{12}$ and $C_{13}$ are linked with the response of uniaxial stress to the strain along a perpendicular axis. $C_{14}$ and $C_{44}$ determine the response of the crystal to shearing stress. Table 2 shows that $C_{66} > C_{44}$. Major implication of this finding is that, the [100] (010) shear should be harder than the [100] (001) shear for $NaSn_2P_2$.

We have illustrated the pressure, $P$, dependent behavior of $C_{ij}$ of $NaSn_2P_2$ in Fig. 2. It is worth noticing that $C_{11}$, $C_{12}$, $C_{14}$, $C_{44}$, and $C_{66}$ show anomalous features at around 1.0 GPa. This is indicative of pressure induced structural instability and can be a subject of further investigation. At $P > 1.5$ GPa, $C_{14}$ assumes small negative values which indicate that slight non-equilibrium internal stress builds up inside the crystal structure as pressure increases.

We have investigated the mechanical stability of $NaSn_2P_2$ at different pressures by employing the well known Born-Huang [39] criteria in their modified form which provides with the necessary and sufficient conditions for mechanical stability for crystals with different symmetries [40]. For the crystal symmetry under consideration, these stability conditions read:

$(C_{11} - C_{12}) > 0$; $C_{44} > 0$; $C_{13}^2 < (½)\{C_{33}(C_{11} + C_{12})\}$;
$C_{14}^2 < (½)\{C_{44}(C_{11} - C_{12})\} \equiv C_{44}C_{66}$ (2)

It is found that all the stability criteria stated above are satisfied for $NaSn_2P_2$ in the pressure range considered here. Once again, there is no reported value of $C_{ij}$ of $NaSn_2P_2$ at any pressure and therefore, the values reported herein cannot be compared.

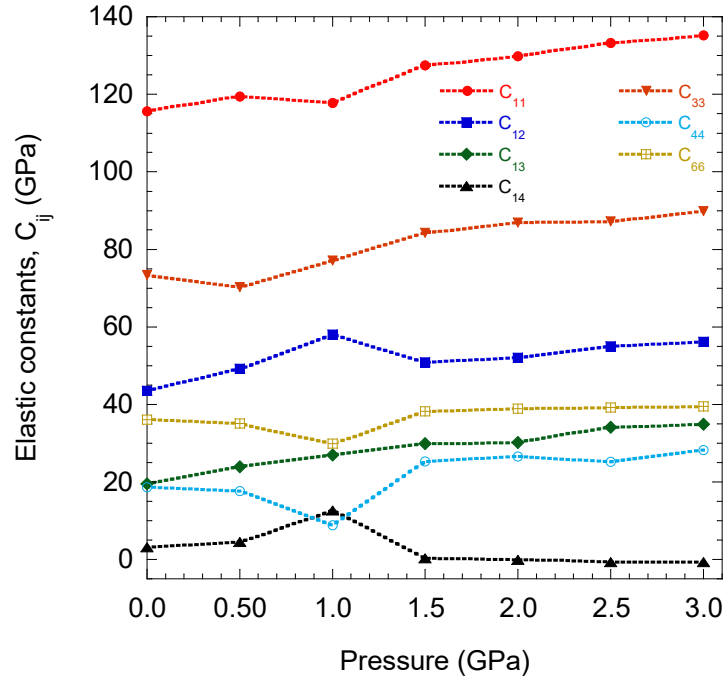

**Fig. 2.** $C_{ij}(P)$ behavior for $NaSn_2P_2$ compound.

The elastic moduli for polycrystalline aggregate and the Poisson's ratio of $NaSn_2P_2$ were calculated from the computed values of $C_{ij}$. Table 3 displays these elastic moduli and the Poisson's ratio of the NaSn$_2$P$_2$ compound at different pressures. As the Voigt approximation [41] asserts, the grain averaged isotropic bulk and shear moduli can be calculated from combinations of various single crystal elastic stiffness constants irrespective of the single crystal structure [41, 42 – 44]. The Voigt approximated bulk and shear moduli are symbolized by $B_V$ and $G_V$, respectively. Reuss [45], on the other hand, proposed different estimates for isotropic bulk and shear moduli from the single crystal elastic constants [42 – 44] assuming an equal stress distributed over the polycrystalline aggregates. The Reuss approximated bulk and shear moduli are denoted here by $B_R$ and $G_R$, respectively. Later on Hill [46] demonstrated that, the Voigt and Reuss approximated values are actually the upper and lower limits of the polycrystalline elastic moduli. A more practical estimate of the bulk and shear moduli are therefore, the arithmetic averages given by, $B = (B_V + B_R)/2$ and $G = (G_V + G_R)/2$. Altogether, this is known as the widely employed VRH approximation. Both Young's modulus, $Y$, and Poisson's ratio, $n$, are linked to the polycrystalline bulk and shear moduli and can be calculated from those [42 – 44]. The Pugh's ratio, expressed as $B/G$, is another useful mechanical indicator and so is the Cauchy pressure ($CP$), defined as $CP = (C_{11} - C_{12})$. We have displayed these two parameters in Table 3 as well.

**Table 3**
Elastic moduli (all in GPa), Pugh's ratio, Poisson's ratio and Cauchy pressure (GPa) of NaSn$_2$P$_2$ at different pressures (GPa).

| $P$ | $B_V$ | $B_R$ | $B$ | $G_V$ | $G_R$ | $G$ | $Y_V$ | $Y_R$ | $Y$ | $B/G$ | $n$ | $CP$ | Ref. |
|---|---|---|---|---|---|---|---|---|---|---|---|---|---|
| 0 | 52.18 | 47.91 | 50.04 | 29.48 | 25.81 | 27.64 | 74.42 | 65.63 | 70.03 | 1.81 | 0.267 | 24.83 | This |
| 0.5 | 55.91 | 50.13 | 53.02 | 28.22 | 24.12 | 26.17 | 72.47 | 62.35 | 67.42 | 2.03 | 0.288 | 31.54 | This |
| 1.0 | 59.59 | 54.45 | 57.03 | 22.90 | 6.41 | 14.66 | 60.91 | 18.51 | 40.50 | 3.89 | 0.382 | 49.15 | This |
| 1.5 | 62.28 | 58.25 | 60.26 | 33.00 | 31.34 | 32.17 | 84.15 | 79.72 | 81.93 | 1.87 | 0.273 | 25.58 | This |
| 2.0 | 63.51 | 59.55 | 61.53 | 34.03 | 32.50 | 33.26 | 86.62 | 82.49 | 84.56 | 1.85 | 0.271 | 25.45 | This |
| 2.5 | 66.68 | 62.28 | 64.48 | 33.28 | 31.45 | 32.37 | 85.59 | 80.77 | 83.18 | 1.99 | 0.285 | 29.83 | This |
| 3.0 | 67.99 | 63.73 | 65.86 | 34.80 | 33.50 | 34.15 | 89.18 | 85.51 | 87.35 | 1.93 | 0.279 | 27.98 | This |

The small values of elastic moduli once again demonstrate that NaSn$_2$P$_2$ is a fairly soft compound. Among the three elastic moduli, the Young's modulus is the largest signifying that the van der Waals-type NaSn$_2$P$_2$ layered pnictide is most resistant against tensile stress. At all pressures considered, $G < B$. This implies that the mechanical failure mode of NaSn$_2$P$_2$ should be controlled by shape deforming stress rather than the volume changing one. The hardness of a solid depends strongly on two parameters, the shear modulus $G$ and the shear related stiffness constant $C_{44}$ [47, 48]. Values of both these parameters are very low (Tables 2 and 3). Therefore, the predicted hardness of NaSn$_2$P$_2$ is very low compared to many other layered ternaries [48]. Particularly noteworthy is the case at a pressure of 1.0 GPa where the values of $C_{44}$ and $G$ are only 8.83 GPa and 14.66 GPa, respectively. These extraordinarily low values should make NaSn$_2$P$_2$ very soft at this particular pressure. The Pugh's ratio, Poisson's ratio and Cauchy pressure are well known indicators used widely to demarcate the solids into brittle or ductile groups where the critical value of $B/G$ ratio is 1.75 [49], $n$ is 0.26 [50, 51] and $CP$ is zero [52]. The values of $B/G$ exceeding 1.75, $n$ higher than 0.26, and positive values of $CP$ signify ductile behavior of solids. The calculated values of all these parameters presented in Table 3 demonstrate that NaSn$_2$P$_2$ is ductile in nature and the degree of ductility is a non-monotonic function of pressure. Once again, we see anomalous behavior at a pressure of 1.0 GPa where the van der Waals-type NaSn$_2$P$_2$ layered pnictide becomes highly ductile and soft. Besides, the magnitude and sign of $CP$ provides us with information regarding the dominant nature of chemical bonding in the compound [52]. We have found high and positive values of $CP$ at all pressures which imply that the material under investigation possesses significant metallic bonding. We infer the ductility of NaSn$_2$P$_2$ to the metallic bonding. Poisson's ratio holds an important part in assessing a number of mechanical properties of crystalline solids. For example, it can assess the stability of solids against shear. Low value of $n$ is indicative of stability against shear [53]. Furthermore, Poisson's ratio is linked to the nature of interatomic forces present in bonding in the solids [54]. In materials with central force interaction dominating, $n$ lies in the range from 0.25 to 0.50 and for non-central force materials, $n$ lies outside of this domain. Considering the calculated values of $n$ at different pressures as displayed in Table 3, it becomes clear that NaSn$_2$P$_2$ is less tolerant against shear and central force interaction dominates the atomic bonding. Both these characteristics become highly amplified at a pressure of 1.0 GPa where the Poisson's ratio becomes anomalously high, 0.382. The findings from the Pugh's ratio, Poisson's ratio and the Cauchy pressure are in complete accord with one another.

As far as possible engineering applications are concerned, machinability index, $\mu_M = B/C_{44}$ [55], is an important performance indicator. The machinability index gives an idea about the degree of ease/difficulty with which a solid can be cut, put into different shapes etc. A higher value of $\mu_M$ indicates greater ease to shape manipulation and also to greater dry lubrication. Such solids require less power to be machined and they can be shaped quickly in to the desired geometry. We have calculated the machinability index of NaSn$_2$P$_2$ at different pressures. The machinability index of NaSn$_2$P$_2$ is high at all pressures. Aluminum (Al) a soft, ductile and machinable element has a $\mu_M \sim 2.0$ [55]. The lowest value of $\mu_M$ of NaSn$_2$P$_2$ is found to be 2.31 at a pressure of 2.0 GPa. The highest value is 6.46 at a pressure of 1.0 GPa. Overall, $\mu_M$ discloses non-monotonic pressure dependence with values of 2.68, 3.00, 6.46, 2.38, 2.31, 2.56 and 2.34, at pressures of 0.0, 0.5, 1.0, 1.5, 2.0, 2.5 and 3.0 GPa respectively.

Almost all of the crystalline solids are elastically anisotropic. Elastic anisotropy often controls the mechanical response of a solid due to variety of external stresses. Therefore, study and understanding of elastic anisotropy is important for materials design, particularly for compounds with layered character. The elastic anisotropy has significant implications in engineering science owing to its correlation with the creation and propagation of microcracks in the crystals, formation of grain boundaries, possible mechanical failure modes etc. We have calculated a number of anisotropy indices for NaSn$_2$P$_2$ at different pressures and have presented those in Table 4. The widely used anisotropy indices are the shear anisotropy factors: $A_1$, $A_2$ and $A_3$, the anisotropy indices for the bulk and shear moduli, $A_B$ and $A_G$, respectively and the universal anisotropy index, $A_U$, applicable for all crystal systems irrespective of their symmetry. The detail formalisms used to calculate these anisotropy indicators can be found elsewhere [56, 57]. Besides, we have calculated the compressibility ratio along $c$- and $a$-direction, $k_c/k_a$, of NaSn$_2$P$_2$ at different representative pressures which is another important anisotropy parameter (Table 4). To visualize the variation of the elastic moduli, compressibility (inverse of bulk modulus) and Poisson's ratio in three-dimensional (3D) space, we have presented the ELATE [58] generated 3D plots in Figs. 3.

**Table 4**
Calculated indices of elastic anisotropy and compressibility ratio at different pressures of layered vdW compound NaSn$_2$P$_2$.

| $P$ (GPa) | $A_1$ | $A_2$ | $A_3$ | $A_B$ | $A_G$ | $A_U$ | $k_c/k_a$ | Ref. |
|---|---|---|---|---|---|---|---|---|
| 0 | 0.4974 | 0.4974 | 1.0 | 4.2662 | 6.6377 | 0.8001 | 2.2277 | This |
| 0.5 | 0.4984 | 0.4984 | 1.0 | 5.4508 | 7.8334 | 0.9652 | 2.6091 | This |
| 1.0 | 0.2504 | 0.2504 | 1.0 | 4.5072 | 6.2607 | 2.9571 | 2.4288 | This |
| 1.5 | 0.6668 | 0.6668 | 1.0 | 3.3436 | 2.5800 | 0.3340 | 2.1825 | This |
| 2.0 | 0.6801 | 0.6801 | 1.0 | 3.2179 | 2.2997 | 0.3019 | 2.1408 | This |
| 2.5 | 0.6605 | 0.6605 | 1.0 | 3.4119 | 2.8271 | 0.3616 | 2.2585 | This |
| 3.0 | 0.7258 | 0.7258 | 1.0 | 3.2341 | 1.9034 | 0.2609 | 2.2120 | This |

It is found that $A_1 = A_2$ and both these shear anisotropy indices depart significantly from unity. $A_3$, on the other hand is one. This is a consequence of trigonal crystal symmetry. The universal anisotropy index deviates strongly from zero. The ratio between the compressibilities is quite large, deviating strongly from unity. Large values of $k_c/k_a$ imply that the crystal is much more compressible in the $c$-direction. Both $A_1$ and $A_2$ increases with pressure except at $P = 1.0$ GPa, where they show an anomalous decrease. Both anisotropy indices for bulk and shear moduli generally decrease with increasing pressure. Exception is found around 1.0 GPa. For perfectly isotropic crystals $A_B = A_G = 0$. Strong deviation from zero implies highly anisotropic nature of NaSn$_2$P$_2$. Like $A_B$ and $A_G$ the universal anisotropy index, $A_U$, also decreases with increasing pressure except at 1.0 GPa, where the index is greatly enhanced. For elastically isotropic crystals, $A_G = 0$. All the different indicators of anisotropy suggest together that the compound under investigation possesses highly anisotropic elastic property and its mechanical properties are predicted to be highly direction dependent.

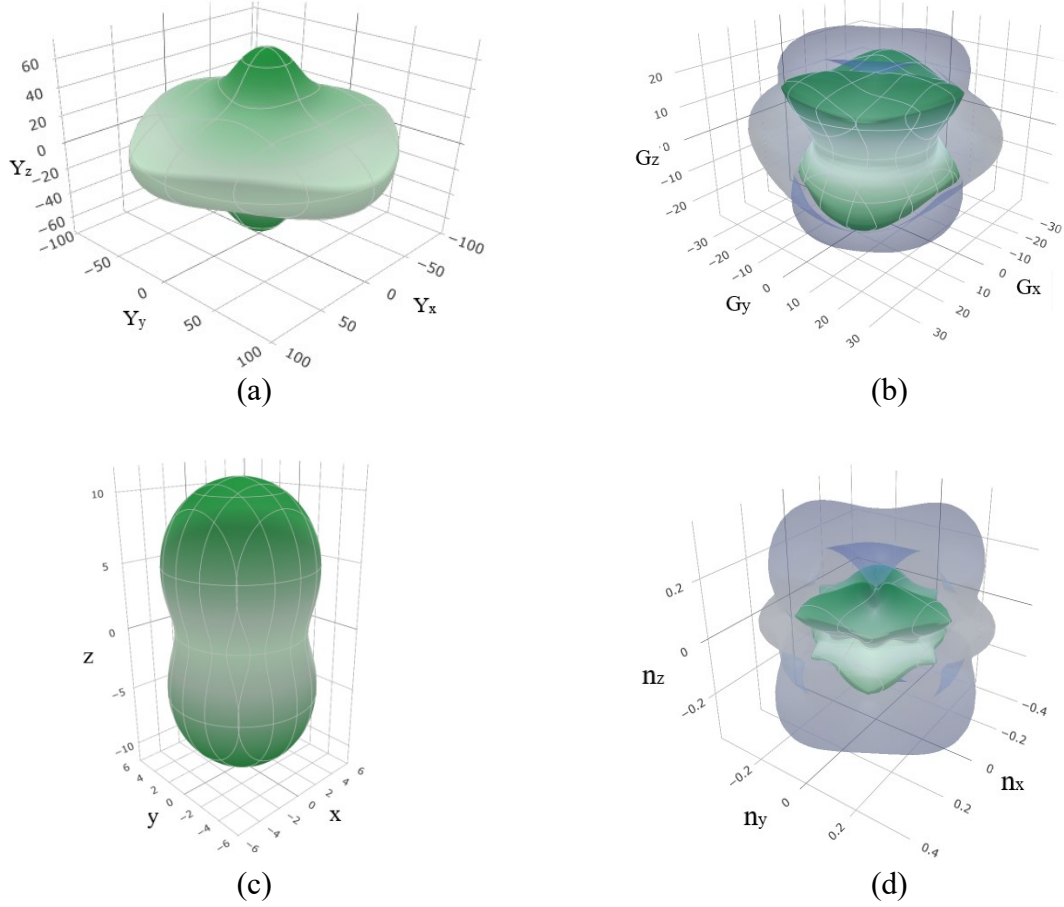

**Fig. 3.** 3D plots of (a) Young's modulus, (b) modulus of rigidity, (c) compressibility and (d) Poisson's ratio of NaSn$_2$P$_2$.

Figs. 3 complement the numerical analysis of the anisotropy indices quite nicely. We can visualize the highly anisotropic nature of the elastic constants and Poisson's ratio from these plots. The 3D plots clearly reveal that NaSn$_2$P$_2$ possesses anisotropic elastic and mechanical properties both within the *ab*-plane and along the out-of-plane direction.

*3.2. Electronic band structure of NaSn$_2$P$_2$*

Almost all of the technologically important physical properties of solids are determined by the valence and conduction electrons in the materials. The behavior of these electrons, on the other hand, is dependent on the nature of their energy dispersion (E(*k*)) within the Brillouin zone. This variation of the electronic energy in different bands with momentum maps the electronic band structure. We have calculated the electronic band structure with the optimized crystal structure of NaSn$_2$P$_2$. The electronic energy dispersion curves under different pressures along the high-symmetry directions of the Brillouin zone of NaSn$_2$P$_2$ are shown in Fig. 4. The total and partial density of states (TDOS and PDOS, respectively), as a function of energy, ($E - E_F$), extracted from the band dispersions at different pressures are disclosed in Figs. 5. The vertical straight line signifies the Fermi energy, $E_F$, which has been set to zero. To elucidate the contribution of each electronic orbital to the TDOS and their relevance in the formation of atomic bonding, we have calculated the orbital resolved PDOS of Na, Sn, and P atoms in $NaSn_2P_2$ (Figs. 5).

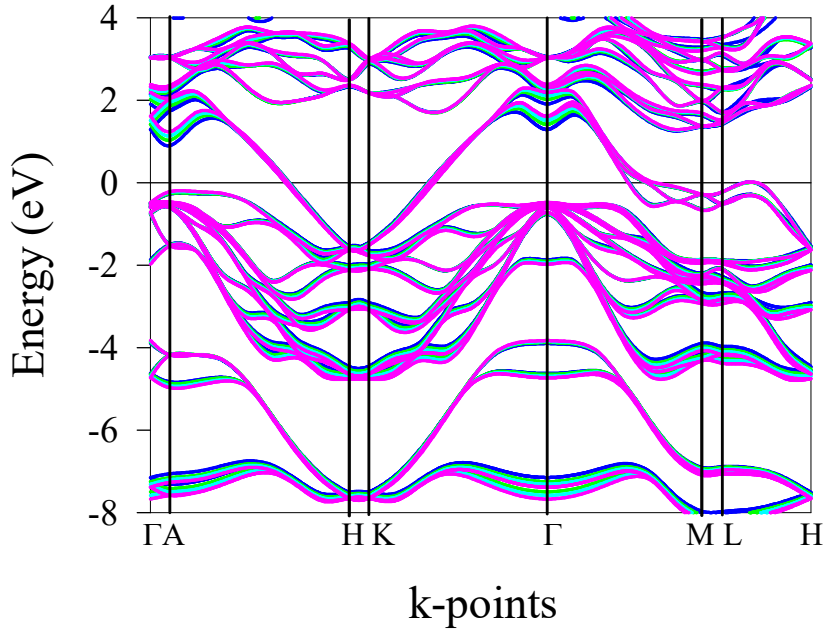

**Fig. 4.** Electronic band structures of $NaSn_2P_2$ along high symmetry directions in the first Brillouin zone (blue, green, cyan and pink colors represent energy dispersion curves for 0.0, 1.0, 2.0 and 3.0 GPa pressures, respectively).

The band structure features clearly disclose that a number of bands with varying degree of dispersion cross the Fermi level. This confirms the metallic character of layered vdW $NaSn_2P_2$ ternary. Highly dispersive bands running along $A – H$ and $K – \Gamma$ directions show electronic character. The charge transport properties are expected to be controlled by these bands. The bands crossing the Fermi level close to the $M$ and $L$ symmetry points are significantly less dispersive. Highly dispersive bands imply low charge carrier effective mass and high charge mobility. The presence of lowly dispersive $E(k)$ around the $M$ and $L$ regions indicate that the material under study possess significant anisotropy in charge transport along different crystallographic directions as well as in the momentum space.

In the preceding section, we have gathered ample indication that $NaSn_2P_2$ is a highly deformable soft compound. Therefore, one expects that application of pressure should change the lattice constants and crystal volumes significantly. In general significant change in the cell volume leads to a notable change in the band structure. This is because the periodicity of the ionic potential changes and affects the band dispersion. Surprisingly, very little change in the $E(k)$ features is observed as a function of pressure (Fig. 4). Possible reason might be that the signs of the pressure coefficient of the deformation potential along different crystallographic axes are different and altogether they tend to cancel each other resulting in a negligible overall effect of pressure on the electronic band structure.

The electronic energy density of states extracted from the band structure calculations are shown in Fig. 5. Like in the electronic band structure results, applied pressure seems to have very little effect on both PDOS and TDOS profiles. Finite TDOS (~ 3.0 states/eV) at the Fermi level confirms metallicity of $NaSn_2P_2$. There is significant hybridization between the Sn $5p$ electronic states and the P $3p$ electronic states near the Fermi energy. The same electronic orbitals constitute the TDOS at the Fermi level, $N(\varepsilon_F)$, almost exclusively, with a small contribution from the Na $3s$ and Sn $5s$ electronic states. Such strong hybridization and enhanced DOS indicate that covalent bonding can form between Sn and P atoms. Compared to the PDOS due to Sn and P, the PDOS due to Na is much smaller both in the valence band below $\varepsilon_F$ and conduction band above it. Small PDOS and relatively weaker hybridization among Na $3s$ and atomic orbitals of Sn and P imply that these atoms are not strongly bonded. The Na atomic layer is somewhat isolated from the other atoms in $NaSn_2P_2$. We predict metallic bonding in these Na layers which should play a significant role in making the overall structure ductile, soft, layered and highly machinable. There are a number of peaks in the TDOS, e.g., at ~ -10.0 eV, -7.0 eV, -5.0 eV, -2.0 eV, -1.0 eV, 2.0 eV and 3.0 eV. All these peaks are expected to play important roles in optical transitions. Particularly, the peaks close the Fermi level should control the charge transport and related electrical properties. The close proximity of the peak at ~ -1.0 eV to the Fermi energy suggests that $NaSn_2P_2$ might be a suitable material for band engineering.

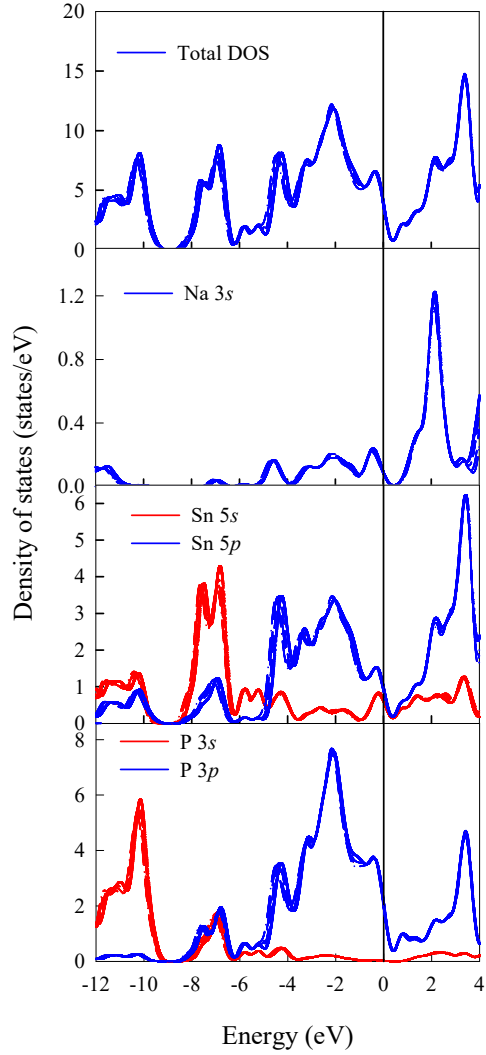

**Fig. 5.** TDOS and PDOS of $NaSn_2P_2$ at different pressures (solid, short-dashed, long-dashed and dotted-dashed lines are for 0.0, 1.0, 2.0 and 3.0 GPa pressures, respectively). The vertical line shows the Fermi energy (set to zero energy).

*3.3. Debye temperature and melting temperature of $NaSn_2P_2$*

Debye temperature, $\theta_D$, is one of the most important thermo-physical parameter of solids. Debye temperature can be defined as the temperature at which all the atomic modes of vibrations become activated. The frequency of the highest energy vibrational mode, $\nu_D$, is known as the Debye frequency. The relationship between $\theta_D$ and $\nu_D$ is established via $h\nu_D = k_B\theta_D$, where $h$ is the Planck constant and $k_B$ denotes the Boltzmann constant. Debye temperature depends on a number of physical parameters; among those the average inter-atomic force plays an important role. In general, the stiffer is the structure, higher is the Debye temperature. Besides, $\theta_D$ is closely related with the melting temperature, phonon specific heat, phonon thermal conductivity, superconducting transition temperature in phonon mediated superconductors, vacancy formation energy in crystals, etc. [59, 60]. There are several approaches of determining the value of $\theta_D$. In this section, we adopt the methodology developed by Anderson [61]. This method yields reliable estimates of the Debye temperature from the elastic moduli of solids and have been used extensively to determine $\theta_D$ for solids with different crystal symmetries and electronic band structures [62 – 67]. The expression for $\theta_D$ due to Anderson [61] reads,

$$\theta_D = \frac{h}{k_B}\left[\left(\frac{3n}{4\pi}\right)\frac{N_A \rho}{M}\right]^{1/3} v_m$$

where $N_A$, $\rho$, $M$ and $n$ denote the Avagadro's number, crystal mass density, molecular mass and the number of atoms in the molecule, respectively. The average sound velocity $v_m$ in the solid is determined from,

$$v_m = \left[\frac{1}{3}\left(\frac{1}{v_l^3} + \frac{2}{v_t^3}\right)\right]^{-1/3}$$

where $v_l$ and $v_t$ are the longitudinal and transverse sound velocities in the compound which can be obtained from the elastic moduli and density as follows,

$$v_l = \left[\frac{3B + 4G}{3\rho}\right]^{1/2}$$

and

$$v_t = \left[\frac{G}{\rho}\right]^{1/2}$$

The pressure dependent values of $\theta_D$ of NaSn$_2$P$_2$ are given in Table 5.

**Table 5**
Density, transverse, longitudinal, mean velocities of sound and Debye temperature for $NaSn_2P_2$ under different pressures.

| $P$ (GPa) | $\rho$ (gm/cc) | $v_t$(km/sec) | $v_l$ (km/sec) | $v_m$(km/sec) | $\theta_D$ (K) | Ref. |
|---|---|---|---|---|---|---|
| 0.0 | 4.819 | 2.3949 | 4.2461 | 2.6636 | 282 | This |
| 0.5 | 4.866 | 2.3191 | 4.2503 | 2.5861 | 274 | This |
| 1.0 | 4.916 | 1.7269 | 3.9466 | 1.9497 | 208 | This |
| 1.5 | 4.965 | 2.5455 | 4.5578 | 2.8330 | 302 | This |
| 2.0 | 5.011 | 2.5762 | 4.5964 | 2.8667 | 307 | This |
| 2.5 | 5.052 | 2.5313 | 4.6157 | 2.8216 | 303 | This |
| 3.0 | 5.096 | 2.5887 | 4.6751 | 2.8834 | 311 | This |

Once again, we see that the Debye temperature shows an anomalous fall at a pressure of 1.0 GPa. At other pressures, $\theta_D$ exhibits an increasing trend with increasing pressure. This is the conventional behavior since pressure makes the crystal stiff and thereby increases the Debye temperature. It is interesting to note that the sound velocities also decrease significantly at 1.0 GPa and the difference between the velocities in the transverse and longitudinal modes also becomes pronounced.

Melting temperature, $T_m$, is another important thermo-physical parameter which limits the application of a material at high temperatures. $T_m$ also reflects the overall bonding strength of solids. We have estimated the melting point of at $NaSn_2P_2$ different pressures following the formalism developed in Ref. [68]. The following expression has been used to estimate $T_m$ (in K) from the calculated elastic constants: $T_m = 354 + \frac{4.5(2C_{11}+C_{33})}{3}$. The pressure variation of $T_m$ is shown in Fig. 6.

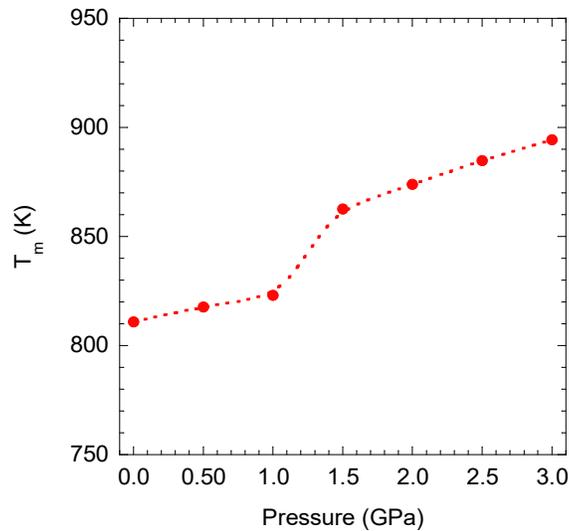

**Fig. 6.** Variation of the melting temperature of $NaSn_2P_2$ with different pressures.

Fig. 6 discloses that $T_m$ is very low for NaSn$_2$P$_2$. For example, the MAX phase ternaries with layered structure have melting temperatures in the range 1400 K – 2300 K [48]. It is also observed that $T_m$ varies non-monotonically around 1.0 GPa. The low values of $T_m$ are expected because of highly soft nature of the structure of layered vdW NaSn$_2$P$_2$ compound.

*3.4. Superconductivity in NaSn$_2$P$_2$ and pressure*

NaSn$_2$P$_2$ exhibits superconductivity at low temperature. For conventional low-$T_c$ superconductors where electron-phonon interactions lead to Cooper pairing, the superconducting transition temperature can be expressed as [69],

$$T_c = \frac{\theta_D}{1.45} \exp\left\{-\frac{1.04(1+\lambda)}{\lambda - \mu^*(1+0.62\lambda)}\right\}$$

where $\lambda$ is the electron phonon coupling constant and $\mu^*$ is the repulsive Coulomb pseudopotential. One can express the coupling constant as, $\lambda = N(E_F)V_{e\text{-ph}}$, with $V_{e\text{-ph}}$ representing the electron-phonon interaction energy responsible for the Fermi surface instability and eventual Cooper pairing. From the TDOS plots in Fig. 5, one observes that there is little variation in the $N(E_F)$ with pressure. Therefore, the pressure induced change in $\lambda$ will be determined by the possible change in $V_{e\text{-ph}}$ with pressure. For a fixed value of $\lambda$, $T_c$ is positively and linearly correlated with $\theta_D$. Except at a pressure of 1.0 GPa, we have found that $\theta_D$ of NaSn$_2$P$_2$ increases systematically with pressure and this should contribute positively to an increase in the $T_c$. The situation at 1.0 GPa is complex. The Debye temperature decreases significantly at this pressure but there are signs of lattice softening as well. Such lattice softening can strongly enhance $\lambda$ via strengthening $V_{e\text{-ph}}$ [70]. Under this situation, the enhanced value of electron-phonon coupling constant can lead to a substantial increase in $T_c$ even though is $\theta_D$ suppressed. It should be noted that the value of $\mu^*$ is not expected to be affected by pressure to a great deal [71, 72].

*3.5. Thermodynamic properties of NaSn$_2$P$_2$*

Compounds are often used at different pressures and temperatures. Therefore, study of the variation of different thermodynamic parameters with pressure and temperature is important in view of potential engineering applications. All the pressure and temperature dependent thermodynamic properties presented in this section are obtained from the quasi-harmonic Debye approximation (QHDA). The QHDA is a lattice dynamical model of solid-state physics used to describe volume-dependent thermal effects. It is based on the assumption that the harmonic approximation holds for every value of the lattice constant, which is to be viewed as an adjustable parameter as a function of pressure and temperature. Details regarding this approximation can be found elsewhere [73]. The validity of QHDA is better at low temperatures; therefore, we have limited our analysis of thermo-physical parameters to temperatures up to 700 K.

Figs. 7a and 7b show the temperature and pressure dependences of bulk modulus of $NaSn_2P_2$. It is observed that at low-$T$, the bulk modulus decreases very slowly with increasing temperature. This is standard behavior related to the third law of thermodynamics. An almost linear decrease of the bulk modulus starts at around 100 K and same trend is maintained up to 700 K (Fig. 7a). The bulk modulus increases quasi-linearly with pressure (Fig. 7b). This is also standard behavior since application of external pressure makes the crystal structure mechanically stronger. The $T$- and $P$-dependent behavior of the Debye temperature is displayed in Figs. 7c and 7d, respectively. From Figs. 7c and 7d, it is seen that the temperature and pressure dependences of the Debye temperature follow the same qualitative trend shown by the bulk modulus. The primary underlying physical reasons for this similarity can be understood as follows: as temperature increases the crystal stiffness decreases leading to a decrease in both bulk modulus and Debye temperature; on the other hand increase in pressure stiffens the crystal resulting in an increase in bulk modulus and Debye temperature. The $T$- and $P$-dependent variation of the volume thermal expansion coefficients (VTEC) are plotted in Fig. 7e and 7f, respectively. The plots are characterized by high values of VTEC at room-temperatures and above. This is consistent with the elastic constant and moduli results for layered vdW $NaSn_2P_2$ compound which clearly disclosed the soft nature of this compound. Otherwise, Fig. 7e and 7f show typical behavior in their variation with respect to temperature and pressure. Temperature and pressure dependent specific heat under constant volume, $C_V$, and constant pressure, $C_P$, are shown in Fig. 7g and 7h, respectively. In the low-$T$ region, both $C_V$ and $C_P$ follow the well-known Debye $T^3$-law. At high temperatures $C_V$ approaches the classical Dulong-Petit limit. At high temperatures $C_P$ is higher than $C_V$. This is somewhat expected in compounds with high volume thermal expansion coefficient. Overall, the temperature and pressure dependences of $C_P$ and $C_V$ are qualitatively similar to those for VTEC. This is natural because VTEC and specific heat are linearly dependent on each other.

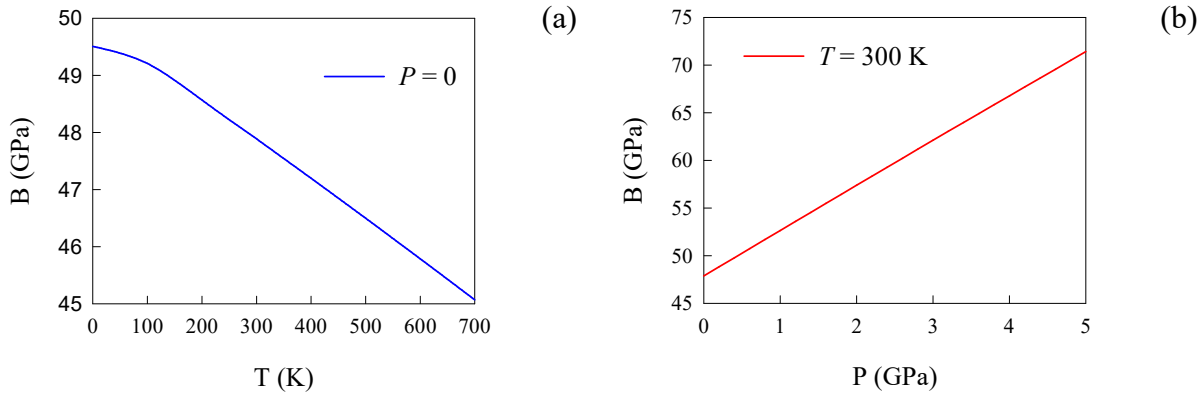

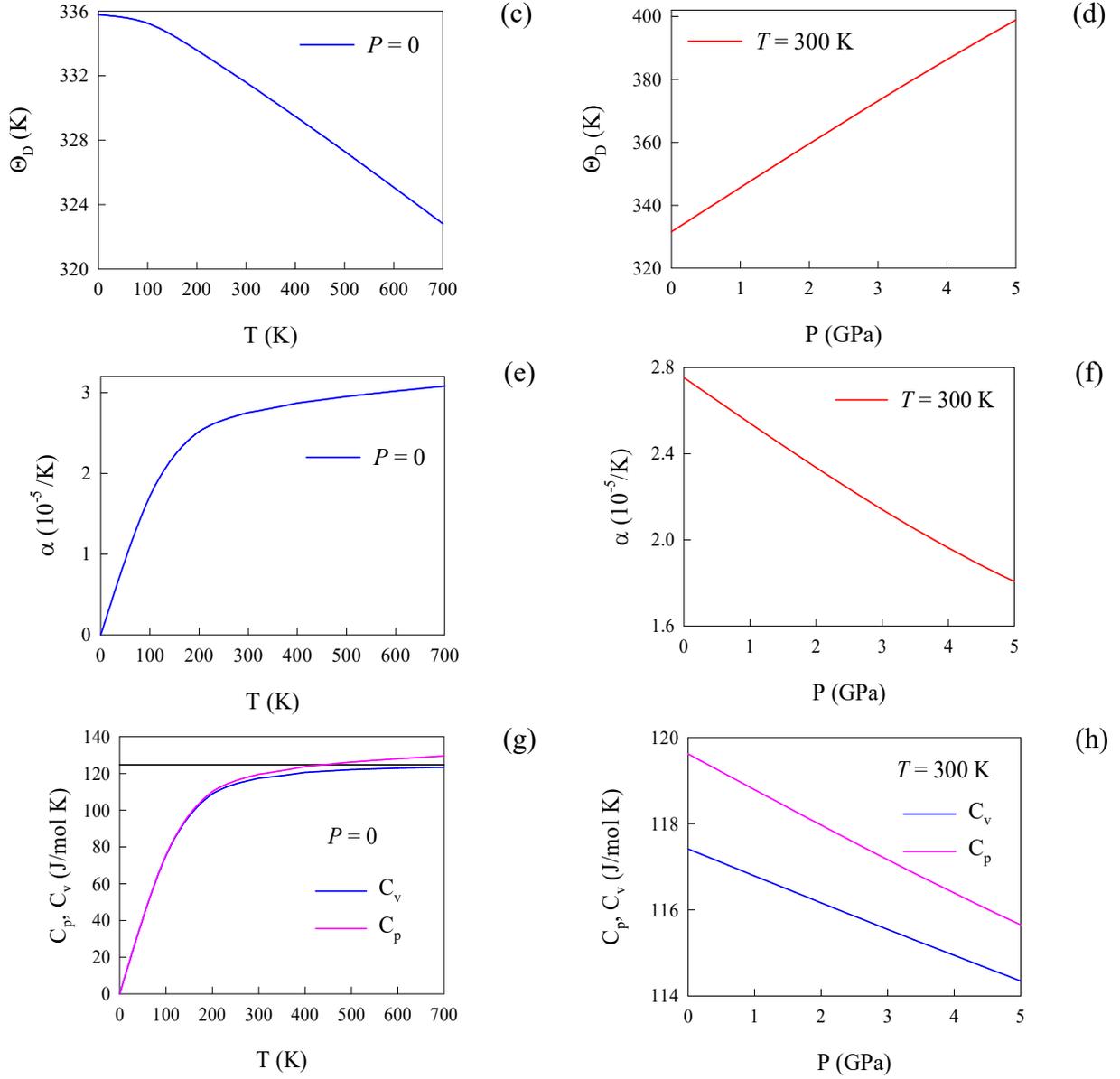

**Fig. 7.** The temperature and pressure dependences of (a, b) bulk modulus, (c, d) Debye temperature, (e, f) VTEC and (g, h) specific heats $C_V$ and $C_P$ of $NaSn_2P_2$. The horizontal line in Fig. 7g marks the classical Dulong-Petit limit of the specific heat.

### 3.6. Optical properties of $NaSn_2P_2$

Exploration of the potential of a newly synthesized compound for various possible applications is incomplete without a thorough understanding of the energy dependent optical parameters. Optical parameters of a compound determine its response to the incident electromagnetic radiation. The optical response to visible part of the electromagnetic spectra is particularly important from the view of optoelectronic device applications. This optical response to the incident light is completely determined by the various energy dependent (or, equivalently, frequency dependent) optical parameters, namely, the real and the imaginary parts of the dielectric constants, $\varepsilon_1(\omega)$ and $\varepsilon_2(\omega)$, respectively, real part of refractive index $n(\omega)$, extinction coefficient $k(\omega)$, real and imaginary parts of the optical conductivity $\sigma_1(\omega)$ and $\sigma_2(\omega)$, respectively, reflectivity $R(\omega)$, absorption coefficient $\alpha(\omega)$, and the loss function $L(\omega)$. The calculated optical parameter spectra for $NaSn_2P_2$ for photon energies up to 20 eV, with polarization vectors along [100] and [001] directions, are displayed in Figs. 8. The electric field polarization dependent response of the optical constants supplies valuable information regarding optical and electronic anisotropy. Study of optical properties yields information regarding photon induced electronic polarizations and electronic transitions between different electronic states. The later is closely linked with the underlying electronic band structure and DOS of the electronic states involved.

Fig. 8a shows the energy dependent behavior of the real and imaginary parts of the dielectric constants. Both $\varepsilon_1(\omega)$ and $\varepsilon_2(\omega)$ show metallic characteristics. In metallic compounds, in the low energy region where $\omega\tau << 1$ ($\tau$ denotes the electronic relaxation time and $\omega$ is the angular frequency), $\varepsilon_2(\omega)$ dominates the optical response since $\varepsilon_2(\omega) \sim \sigma_1(\omega)/\omega$. One the other hand, $\sigma_1(\omega)$ is finite at low frequencies in metallic compounds. In the high energy limit, $\omega\tau >> 1$ (high frequency region), the real part approaches unity and the imaginary part becomes very small. This high energy region is characterized by the inductive nature of electronic motion induced by high-frequency of the incident electromagnetic radiation. It is observed from Fig. 8a that $\varepsilon_1$ crosses zero from below at ~ 18.0 eV and at the same energy $\varepsilon_2$ flattens to a very low value irrespective of the electric field polarization of the incident electromagnetic wave (EMW). Consequently, $NaSn_2P_2$ is predicted to become transparent to incident electromagnetic radiation above 18.0 eV. The peaks observed in $\varepsilon_1(\omega)$ and $\varepsilon_2(\omega)$ are due to high values of the product of density of states of the electronic orbitals involved in the process of absorption of EMW. To be specific, the peaks in $\varepsilon_1(\omega)$ and $\varepsilon_2(\omega)$ seen at around energies of 3.0 – 4.5 eV originate due to optical transitions of electrons residing in the high DOS region at ~ -1.0 eV below the Fermi level to the high DOS region at energies around 2.0 – 3.0 eV above the Fermi level. Both $\varepsilon_1(\omega)$ and $\varepsilon_2(\omega)$ show moderate optical anisotropy, particularly at low energies. At high energies of the EMW the anisotropy diminishes greatly.

Fig. 8b displays the reflectivity spectrum of $NaSn_2P_2$. The compound under investigation is a very efficient reflector of infrared and near visible light. The reflectivity falls just below 40% for a large part of the visible spectrum. $R(\omega)$ rises again in the ultraviolet (UV) region and sustains high values over a significant portion of the spectrum. The polarization dependent behavior of $R(\omega)$ is rather weak. $R(\omega)$ falls sharply at ~ 18.0 eV as the material becomes transparent to the incident EMW.

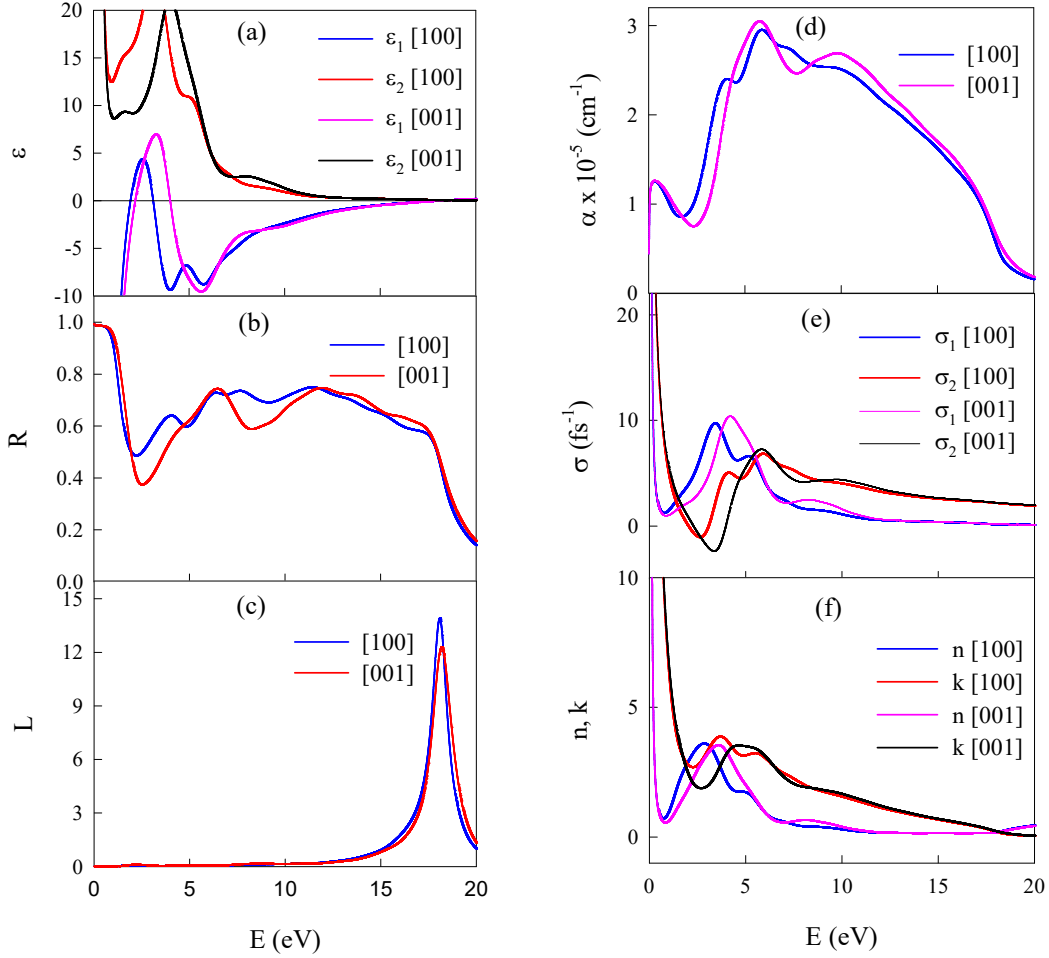

**Fig. 8.** The energy (or, equivalently, frequency) dependent (a) dielectric constant, (b) reflectivity, (c) loss function, (d) absorption coefficient, (e) optical conductivity, and (f) refractive index of $NaSn_2P_2$ with electric field polarization vectors along [100] and [001] directions.

Fig. 8c shows the loss function profile, $L(\omega)$, of $NaSn_2P_2$. The loss function measures the response of the electronic system when high energy electrons traverse the material. Under such circumstances the energy of the fast moving charge is attenuated and collective electronic excitations are created, known as the plasma excitations. The peak in the loss function appears precisely when the energy matches the frequency, $\omega_p$, of the plasma excitation. It is seen from Fig. 8c that the plasma frequency of $NaSn_2P_2$ corresponds to photon energy of ~ 18.0 eV. Above the plasma energy, a metal changes its optical behavior to dielectric like response.

Optical absorption coefficient is an important parameter for optoelectronic device applications, particularly in the photovoltaic sector. Fig. 8d shows the absorption spectra of $NaSn_2P_2$. The absorption spectra show metallic characteristics as there is significant photon absorption at very low energies. There is a dip in the absorption at ~ 2.5 eV, $\alpha(\omega)$ starts increasing at higher energies and reaches a peak ~ 6.0 eV. $\alpha(\omega)$ has high values in the energy range from ~ 5.0 eV to 13.0 eV, covering a wide band in the UV region. The absorption coefficient falls sharply at the plasma energy as at this energy the compound becomes largely transparent to incident EMW.

The real and imaginary parts of the optical conductivity are shown in Fig. 8e. Both real and imaginary parts of the optical conductivity are finite and high at zero frequency. This is a hall mark of strong metallic conductivity of NaSn$_2$P$_2$, entirely consistent with the optical spectra for absorption coefficient, dielectric constant and electronic band structure calculations. The optical conductivity decreases with increasing energy in the mid and far UV region. This behavior is typical to metallic systems.

Finally, we have displayed the real and imaginary parts of the complex refractive index in Fig. 8f. The energy dependent refractive index is of critical significance for photonic device fabrication, such as for constructing optical wave-guides. The real part of the complex refractive index, $n(\omega)$, determines the phase velocity of the EMW in the material, while the imaginary part, $k(\omega)$, known as the extinction coefficient measures the amount attenuation of EMW when it travels through the material. The optically dispersive character of a compound is completely determined by the frequency dependent refractive index. Physically, both the real and imaginary parts of the refractive index are linked to the dielectric constant through the relations, $\varepsilon_1(\omega) = n^2(\omega) - k^2(\omega)$ and $\varepsilon_2(\omega) = 2n(\omega)k(\omega)$. These two equations force both real and imaginary parts of the refractive index to assume very low values at high energies as seen in Fig. 8f.

The optical parameters shown in this section exhibit small to moderate level of optical anisotropies with respect to the polarization direction of the incident EMW. Moreover, the level of anisotropy is comparatively higher at low energies.

## 4. Discussion and Conclusions

First-principles density functional theory based calculations have been carried out to study the structural, elastic, electronic, thermodynamic, and optical properties of recently discovered layered ternary vdW NaSn$_2$P$_2$ superconductor in the trigonal crystal symmetry. Various pressure and temperature dependent thermo-physical parameters have been explored using the QHDA for the first time. The frequency/energy dependent spectra of optical parameters have also been estimated for different electric field polarizations of the EMW for the first time. The optimized crystal structure and ground state lattice parameters agree reasonably well with earlier studies [4]. The observed deviation between the theoretically estimated lattice parameters and experimentally found ones can partly be accounted for by considering the high value of the volume thermal expansion coefficient of NaSn$_2$P$_2$ as found in Section 3.5. The magnitudes of various elastic stiffness constants and moduli indicate that NaSn$_2$P$_2$ is a soft material with ductile character. The compound possesses high level of elastic anisotropy consistent with its layered character. The machinability index of NaSn$_2$P$_2$ is very high emphasizing its highly machinable character and possibility of application as dry lubricant. The Cauchy pressure asserts that central force should dominate in atomic bondings. Both the Pugh and Poisson's ratios indicate that significant metallic bonding should be present in $NaSn_2P_2$. The Debye temperature was calculated from the elastic constants and from the application of QHDA. Values of $\theta_D$ obtained from these two methods agree quite well with each other. Low value of $\theta_D$ implies softness of $NaSn_2P_2$, complementing the results obtained from the analysis of the elastic parameters. The melting temperatures of $NaSn_2P_2$ at different pressures are low compared to many other layered metallic ternaries. The pressure dependent variation of different elastic constants, moduli, ratios and anisotropy indices are non-monotonous and large anomaly is found around the pressure of 1.0 GPa. The pressure dependence of the bulk modulus investigated via QHDA shows good correspondence with the pressure dependence of the same parameter studied with the help of the stress-strain methodology in-built in the CASTEP code. This provides us with an independent check of the validity of the methodologies employed in this work.

The electronic band structure and DOS studies reveal clear metallic character of $NaSn_2P_2$. Quite surprisingly, little effect of pressure was found in the electronic band structure. The TDOS at the Fermi level was largely insensitive to pressure. Therefore, we predict that the pressure dependent superconducting transition temperature of $NaSn_2P_2$ should be affected mainly by the pressure dependent variation in the Debye temperature. At a pressure of 1.0 GPa, there is possibility of significant softening of the phonon modes which may lead to a large enhancement of the electron phonon coupling constant. It is instructive to note that the Fermi level is located at a steeply falling part of the TDOS profile (Fig. 5). Therefore, a small shift of the DOS profile to higher energies can lead to a significant change in the TDOS at the Fermi level and can induce a large increase in the superconducting $T_c$. Such a shift in energy might be accomplished by doping the compound with suitable dopants. At this point we would like to mention that the value of the experimentally determined Debye temperature from the specific heat measurement of $NaSn_2P_2$ and the theoretically computed PDOS of the same compound [4] show reasonable agreements with the results obtained herein.

The temperature and pressure dependent features of the thermodynamic parameters of $NaSn_2P_2$ compound presented in Section 3.5 exhibit standard behaviors [73]. The QHDA results are systematic and do not show any anomaly in the pressure dependent behavior of the Debye temperature, heat capacities, bulk modulus and VTEC.

Compared to the structural and elastic anisotropies, the anisotropy in the optical parameters is low. The energy dependent behaviors of dielectric constant, absorption coefficient and photoconductivity display metallic character. Peaks in the optical parameters spectra at different energies are consistent with electronic band structure calculations and energy dependent DOS profile. The reflectivity of $NaSn_2P_2$ illustrates that this compound should be a very efficient reflector of infrared and near-visible EMW. Such materials with high reflectivity in the infrared region can be used to confine thermal energy in a specified volume and to minimize heat loss to the environment. Reflectivity is also high over a wide energy band in the UV part of the optical spectra. Moreover, the reflectivity spectrum is almost non-selective in this energy range (3.0 eV to 16.0 eV). Furthermore, the absorption coefficient is high in the near- and mid-ultraviolet regions. Compounds having high absorbance and reflectivity in the ultraviolet can be used as coating materials to protect systems which are prone to photo disintegration or photo-oxidation due to direct exposure to ultraviolet radiation. All these features can be useful for potential optical applications.

To summarize, we have presented a detailed study of structural, elastic, electronic, thermodynamic and optical properties of recently synthesized layered vdW ternary compound $NaSn_2P_2$. Effects of pressure on various physical properties have been investigated. The compound under study possesses a number of interesting features which have been discussed. We hope that this study will inspire researchers to investigate this interesting material in greater details both theoretically and experimentally.

## Acknowledgements

S.H.N. acknowledges the research grant (1151/5/52/RU/Science-07/19-20) from the Faculty of Science, University of Rajshahi, Bangladesh, which partly supported this work.

## Data availability

The data sets generated and/or analyzed in this study are available from the corresponding author on reasonable request.

## List of references

[1]  S. Das, J.A. Robinson, M. Dubey, H. Terrones, M. Terrones, Beyond Graphene: Progress in Novel Two-Dimensional Materials and van der Waals Solids. Annu. Rev. Mater. Res. 45 (2015) 1-27. https://doi.org/10.1146/annurev-matsci-070214-021034.
[2]  I.V. Grigorieva , A.K. Geim, Van der Waals heterostructures.  Nature, 499 (2013) 419-425.
[3]  Y. Goto, A. Yamada, T.D. Matsuda, Y. Aoki, Y. Mizuguchi, SnAs-based layered superconductor $NaSn_2As_2$, J. Phys. Soc. Japan 86 (2017) 123701. https://doi.org/10.7566/JPSJ.86.123701.
[4]  Y. Goto, A. Miura, C. Moriyoshi, Y. Kuroiwa, T.D. Matsuda, Y. Aoki, Y. Mizuguchi, $Na_{1-x}Sn_2P_2$ as a new member of van der Waals-type layered tin pnictide superconductors, Sci. Rep. 8 (2018) 12852. DOI:10.1038/s41598-018-31295-8
[5]  H. Yuwen, Y. Goto, R. Jha, A. Miura, C. Moriyoshi, Y. Kuroiwa, T.D. Matsuda, Y. Aoki, Y. Mizuguchi, Enhanced superconductivity by Na doping in SnAs-based layered compound $Na_{1+x}Sn_{2-x}As_2$, Jpn. J. Appl. Phys. 58 (2019) 083001. https://doi.org/10.7567/1347-4065/ab2eb1.
[6]  P. W. Anderson, The Theory of Superconductivity in the High-$T_c$ Cuprates (1997), Princeton University Press, Princeton, New Jersey.
[7]  X. Xi *et al*., Ising pairing in superconducting $NbSe_2$ atomic layers. Nat. Phys. 12 (2016)139 (2016)
[8]  S.H. Naqib, J.R.Cooper, J.L. Tallon, R.S. Islam, R.A. Chakalov, The Doping Phase Diagram of $Y_{1-x}Ca_xBa_2(Cu_{1-y}Zn_y)_3O_{7-\delta}$ from Transport Measurements: Tracking the Pseudogap below $T_c$. Phys. Rev. B 71 (2005) 054502. https://doi.org/10.1103/PhysRevB.71.054502 .
[9]  Y. Kamihara, T. Watanabe, M. Hirano, H. Hosono, Kamihara, Iron-based layered superconductor $La[O_{1-x}F_x]FeAs$ (*x* = 0.05 – 0.12) with $T_c$ = 26 K. J. Am. Chem. Soc. 130 (2008) 3296. https://doi.org/10.1021/ja800073m.
[10]  M.Q. Arguilla, J. Katoch, K. Krymowski, N.D. Cultrara, J. Xu, X. Xi, A. Hanks, S. Jiang, R.D. Ross, R.J. Koch, S. Ulstrup, A. Bostwick, C. Jozwiak, D. McComb, E. Rotenberg, J. Shan, W. Windl, R.K. Kawakami, J.E. Goldberger, $NaSn_2As_2$: an exfoliatable layered van der Waals Zintl phase, ACS Nano10 (2016) 9500.

https://doi.org/10.1021/acsnano.6b04609.

[11] M.Q. Arguilla, N.D. Cultrara, Z.J. Baum, S. Jiang, R.D. Ross J.E. Goldberger, EuSn$_2$As$_2$: an exfoliatable magnetic layered Zintl–Klemm phase, Inorg. Chem. Front.2 (2017) 378. https://doi.org/10.1039/C6QI00476H.

[12] Kovnir, K. *et al*. Sn$_4$As$_3$ revisited: solvothermal synthesis and crystal and electronic structure, J. Solid State Chem. 182 (2009) 630. https://doi.org/10.1016/j.jssc.2008.12.007.

[13] O. Olofsson, X-Ray investigations of the tin-phosphorus system, Acta Chem. Scand.24 (1970) 1153.

[14] B. Eisenmann, J. Klein, Zintl-Phasenmit Schichtanionen: Darstellung und Kristallstrukturen der isotypenVerbindungen SrSn$_2$As$_2$ und Sr$_{0,87}$Ba$_{0,13}$Sn$_2$As$_2$ sowieeine Einkristallstrukturbestimmung von KSnSb, Z. Anorg. Allg. Chem. 598 (1991) 93. https://doi.org/10.1002/zaac.19915980109.

[15] Z. Lin, G. Wang, C. Le, H. Zhao, N. Liu, J. Hu, L. Guo, X. Chen, Thermal conductivities in NaSnAs, NaSnP, and NaSn$_2$As$_2$: effect of double lone-pair electrons. Phys. Rev. B 95 (2017) 165201.
http:// DOI: 10.1103/PhysRevB.95.165201.

[16] M.Q. Arguilla, N.D. Cultrara, M.R. Scudder, S. Jiang, R.D. Ross, J.E. Goldberger, Optical properties and Raman-active phonon modes of two-dimensional honeycomb Zintl phases, J. Mater. Chem. C5 (2017) 11259. https://doi.org/10.1039/C7TC01907F.

[17] B. Eisenmann, U. Rößler, Crystal structure of sodium phosphidostaimate (II), NaSnP. Zeitschrift für Krist. 213 (1998) 28. https://doi.org/10.1524/ncrs.1998.213.14.28.

[18] K.H. Lii, R.C. Haushalter, Puckered hexagonal nets in 2∞[Sn$_{33}$As$^-_{33}$] and 2∞[Sn$_{33}$Sb$^-_{33}$], J. Solid State Chem. 67 (1987) 374. https://doi.org/10.1016/0022-4596(87)90378-1.

[19] M. Asbrand, B. Eisenmann, Arsenidostannatemitarsen-analogen [SnAsI-schichten: darstellung und struktur von Na[Sn$_2$As$_2$], Na$_{0.3}$Ca$_{0.7}$[Sn$_2$As$_2$], Na$_{0.4}$Sr$_{0.6}$[Sn$_2$As$_2$], Na$_{0.6}$Ba$_{0.4}$[Sn$_2$As$_2$] und K$_{0.3}$Sr$_{0.7}$Sn$_2$As$_2$. Z. Anorg. Allg. Chem. 621 (1995) 576. https://doi.org/10.1002/zaac.19956210415.

[20] P.C. Schmidt, D. Stahl, B. Eisenmann, R. Kniep, V. Eyert, J. Kübler, Electronic structure of the layered compounds K[SnSb], K[SnAs] and Sr[Sn$_2$As$_2$], J. Solid State Chem. 97 (1992) 93. https://doi.org/10.1016/0022-4596(92)90013-L.

[21] M. Asbrand, F.J. Berry, B. Eisenmann, R. Kniep, L.E. Smart, R.C. Thied, Bonding in some Zintl phases: a study by tin-119 Mossbauer spectroscopy, J. Solid State Chem. 118 (1995) 397. https://doi.org/10.1006/jssc.1995.1360.

[22] S. Geller, G.W. Hull, Superconductivity of intermetallic compounds with NaCl-type and related structures. Phys. Rev. Lett. 13 (1964) 127. https://doi.org/10.1103/PhysRevLett.13.127.

[23] M.H. Maareen, On the superconductivity, carrier concentration and the ionic model of Sn$_4$P$_3$ and Sn$_4$As$_3$. Phys. Lett. 29 A (1969) 293. https://doi.org/10.1016/0375-9601(69)90130-3.

[24] S. Huang, H.J. Liu, D.D. Fan, P.H. Jiang, J.H. Liang, G.H. Cao, R.Z. Liang, J. Shi, First-Principles Study of the Thermoelectric Properties of the Zintl Compound KSnSb. J. Phys. Chem. C122 (8) (2018) 4217–4223. https://doi.org/10.1021/acs.jpcc.8b00099.

[25] Saikat Mukhopadhyay, David J. Singh, Thomas L. Reinecke, Ultralow Thermal Conductivity in Cs–Sb–Se Compounds: Lattice Instability versus Lone-Pair Electrons. Chem. Mater. (2020). https://doi.org/10.1021/acs.chemmater.0c02688.

[26] M.M. Mridha, S.H. Naqib, Pressure dependent elastic, electronic, superconducting, and optical properties of ternary barium phosphides (Ba$M$$_2$P$_2$; $M$ = Ni, Rh): DFT based insights, Phys. Scr. 95 (2020) 105809.

[27] M.K. Wu, J.R. Ashburn, C.J. Torng, P.H. Hor, R.L. Meng, L. Gao, Z.J. Huang, Y.Q. Wang, C.W. Chu, Superconductivity at 93 K in a new mixed-phase Y-Ba-Cu-O compound system at ambient pressure, Phys. Rev. Lett. 58 (1987) 908. https://doi.org/10.1103/PhysRevLett.58.908.

[28] W. Kohn, L.J. Sham, Self-consistent equations including exchange and correlation effects, Phys. Rev. 140 (1965) A1133. https://doi.org/10.1103/PhysRev.140.A1133.

[29] S.J. Clark, M.D. Segall, C.J. Pickard, P.J. Hasnip, M.I.J. Probert, K. Refson, M.C. Payne, First principles methods using CASTEP, Z. Kristallogr. 220 (2005) 567-570. https://doi.org/10.1524/zkri.220.5.567.65075.

[30] D. Vanderbilt, Soft self-consistent pseudopotentials in a generalized eigenvalue formalism, Phys. Rev. B 41 (1990) 7892. https://doi.org/10.1103/PhysRevB.41.7892.

[31] T.H. Fischer, J. Almlof, General methods for geometry and wave function optimization, J. Phys. Chem. 96 (1992) 9768. https://doi.org/10.1021/j100203a036.

[32] H.J. Monkhorst, J.D. Pack, Special points for Brillouin-zone integrations, Phys. Rev. B 13 (1976) 5188. https://doi.org/10.1103/PhysRevB.13.5188.

[33] A. Chowdhury, M.A. Ali, M.M. Hossain, M.M. Uddin, S.H. Naqib, A.K.M.A. Islam, Predicted MAX phase Sc$_2$InC: Dynamical stability, vibrational and optical properties, Phys. Status Solidi (b) 255(3)

(2018)1700235. https://doi.org/10.1002/pssb.201700235.

[34] M.A. Ali, M.T. Nasir, M.R. Khatun, A.K.M.A. Islam, S.H. Naqib, An *ab initio* investigation of vibrational, thermodynamic, and optical properties of $Sc_2AlC$ MAX compound, Chin. Phys. B 25(10) (2016) 103102. DOI: 10.1088/1674-1056/25/10/103102.

[35] M. Roknuzzaman, M.A. Hadi, M.J. Abden, M.T. Nasir, A.K.M.A. Islam, M.S. Ali, K. Ostrikov, S.H. Naqib, Physical properties of predicted $Ti_2CdN$ versus existing $Ti_2CdC$ MAX phase: An *ab initio* study, Comput. Mater. Sci. 113 (2016) 148. DOI: 10.1016/j.commatsci.2015.11.039.

[36] M.M. Hossain, S.H. Naqib, Structural, elastic, electronic, and optical properties of layered TiNX (X = F, Cl, Br, I) compounds: a density functional theory study, Mol. Phys. 118 (2020) e1609706. https://doi.org/10.1080/00268976.2019.1609706.

[37] F. Parvin, S.H. Naqib, Structural, elastic, electronic, thermodynamic, and optical properties of layered $BaPd_2As_2$ pnictide superconductor: A first principles investigation, J. Alloys Compd. 780 (2019) 452–460. https://doi.org/10.1016/j.jallcom.2018.12.021.

[38] F. Birch, Finite strain isotherm and velocities for single-crystal and polycrystalline NaCl at high pressures and 300 K, J. Geophys. Res.: Solid Earth 83 (1978) 1257-1268. https://doi.org/10.1029/JB083iB03p01257.

[39] M. Born, K. Huang, Dynamical Theory of Crystal Lattices, Oxford University Press, UK, 1998.

[40] F. Mouhat, F-X. Coudert, Necessary and sufficient elastic stability conditions in various crystal systems, Phys. Rev. B 90, 224104 (2014). https://doi.org/10.1103/PhysRevB.90.224104.

[41] W. Voigt, Lehrbuch der Kristallphysik, Teubner, Leipzig, 1928.

[42] M.A. Hadi, S.H. Naqib, S.R.G. Christopoulos, A. Chroneos, A.K.M.A. Islam, Mechanical behavior, bonding nature and defect processes of $Mo_2ScAlC_2$: A new ordered MAX phase, J. Alloys Compd. 724 (2017) 1167. https://doi.org/10.1016/j.jallcom.2017.07.110.

[43] M.A. Ali, M.M. Hossain, M.A. Hossain, M.T. Nasir, M.M. Uddin, M.Z. Hasan, A.K.M.A. Islam, S.H. Naqib, Recently synthesized $(Zr_{1-x}Ti_x)_2AlC$ ($0 \leq x \leq 1$) solid solutions: Theoretical study of the effects of M mixing on physical properties, J. Alloys Compd. 743 (2018) 146. https://doi.org/10.1016/j.jallcom.2018.01.396.

[44] M.A. Hadi, M. Roknuzzaman, A. Chroneos, S.H. Naqib, A.K.M.A. Islam, R.V. Vovk, K. Ostrikov, Elastic and thermodynamic properties of new $(Zr_{3-x}Ti_x)AlC_2$ MAX-phase solid solutions, Comput. Mater. Sci. 137 (2017) 318. https://doi.org/10.1016/j.commatsci.2017.06.007.

[45] A. Reuss, Berechnung der Fliessgrenze von Mischkristallen auf Grund der PlastizitätsbedingungfürEinkristalle, Z. Angew. Math. Mech. 9 (1929) 49-58. http://dx.doi.org/10.1002/zamm.19290090104.

[46] R. Hill, The elastic behavior of a crystalline aggregate, Proc. Phys. Soc. London, Sect. A 65 (1952) 349. https://doi.org/10.1088/0370-1298/65/5/307.

[47] J.S. Tse, Intrinsic hardness of crystalline solids, J. Superhard Mater. 32 (2010) 177–191. https://doi.org/10.3103/S1063457610030044.

[48] M.A. Ali, M.M. Hossain, A.K.M.A. Islam, S.H. Naqib, Recently predicted ternary boride $Hf_3PB_4$: Insights into the physical properties of this hardest possible boride MAX phase, arXiv:2009.05707.

[49] S.F. Pugh, XCII. Relations between the elastic moduli and the plastic properties of polycrystalline pure metals, Philos. Mag. A 45 (1954) 823-843. https://doi.org/10.4236/jamp.2017.51004.

[50] I.N. Frantsevich, F.F. Voronov, S.A. Bokuta, Elastic constants and elastic moduli of metals and insulators handbook, NaukovaDumka, Kiev, 1983, pp. 60-180.

[51] G. Vaitheeswaran, V. Kanchana, A. Svane, A. Delin, Elastic properties of $MgCNi_3$ - a superconducting perovskite, J. Phys. Condens. Matter 19 (2007) 326214. https://doi.org/10.1088/0953-8984/19/32/326214.

[52] M.E. Eberhart, T.E. Jones, Cauchy pressure and the generalized bonding model for nonmagnetic bcc transition metals, Phys. Rev. B 86 (2012) 134106 and references therein. https://doi.org/10.1103/PhysRevB.86.134106.

[53] P. Ravindran, L. Fast, P.A. Korzhavyi, B. Johansson, J. Wills, O. Eriksson, Density functional theory for calculation of elastic properties of orthorhombic crystals: Application to $TiSi_2$, J. Appl. Phys. 84 (1998) 4891. https://doi.org/10.1063/1.368733.

[54] O.L. Anderson, H.H. Demarest Jr., Elastic constants of the central force model for cubic structures: Polycrystalline aggregates and instabilities, J. Geophys. Res. 76 (1971) 1349. https://doi.org/10.1029/JB076i005p01349.

[55] Z. Sun, D. Music, R. Ahuja, J.M. Schneider, Theoretical investigation of the

| | |
|---|---|
| | bonding and elastic properties of nanolayered ternary nitrides, Phys. Rev. B 71 (2005) 193402. https://doi.org/10.1103/PhysRevB.71.193402. |
| [56] | H.M. Ledbetter, Elastic properties of zinc: A compilation and a review, J. Phys. Chem. Ref. Data. 6 (1977) 1181–1203. https://doi.org/10.1063/1.555564. |
| [57] | S.I. Ranganathan, M. Ostoja-Starzewsk, Universal elastic anisotropy index, Phys. Rev. Lett. 101 (2008) 055504. DOI: 10.1103/PhysRevLett.101.055504. |
| [58] | R. Gaillac, P. Pullumbi, F.-X. Coudert, ELATE: an open-source online application for analysis and visualization of elastic tensors, J. Phys. Condens. Matter 28 (2016) 275201. https://doi.org/10.1088/0953-8984/28/27/275201. |
| [59] | G.Grimvall, S. Sjodin, Correlation of properties of materials to Debye and melting temperatures, Phys. Scr. 10 (1974) 340-352. https://doi.org/10.1088/0031-8949/10/6/011. |
| [60] | J. Bardeen, L.N. Cooper, J.R. Schrieffer, Theory of Superconductivity, Phys. Rev. 108 (1957) 1175. https://doi.org/10.1103/PhysRev.108.1175. |
| [61] | O.L. Anderson, A simplified method for calculating the Debye temperature from elastic constants, J. Phys. Chem. Solids 24 (1963) 909-917. https://doi.org/10.1016/0022-3697(63)90067-2. |
| [62] | F. Parvin, S.H. Naqib, Elastic, thermodynamic, electronic, and optical properties of recently discovered superconducting transition metal boride NbRuB: An *ab-initio* investigation, Chin. Phys. B 26(10) (2017) 106201. DOI: 10.1088/1674-1056/26/10/106201. |
| [63] | M.A. Ali, S.H. Naqib, Recently synthesized $(Ti_{1-x}Mo_x)_2AlC$ ($0 \leq x \leq 0.20$) solid solutions: deciphering the structural, electronic, mechanical and thermodynamic properties via *ab initio* simulations, RSC Adv. 10 (2020) 31535-31546. DOI: 10.1039/d0ra06435a. |
| [64] | M.A. Hadi, S.-R.G. Christopoulos, A. Chroneos, S.H. Naqib, A.K.M.A. Islam, Elastic behaviour and radiation tolerance in Nb-based 211 MAX phases, Mat. Today Commun. 25 (2020) 101499. https://doi.org/10.1016/j.mtcomm.2020.101499. |
| [65] | B.R. Rano, I.M. Syed, S.H. Naqib, Ab initio approach to the elastic, electronic, and optical properties of $MoTe_2$: A topological Weyl semimetal, J. Alloys Compd. 829 (2020) 154522. https://doi.org/10.1016/j.jallcom.2020.154522. |
| [66] | M.I. Naher, S.H. Naqib, Structural, elastic, electronic, bonding, and optical properties of topological $CaSn_3$ semimetal, J. Alloys Compd. 829 (2020) 154509. https://doi.org/10.1016/j.jallcom.2020.154509. |
| [67] | M.A. Afzal, S.H. Naqib, A DFT based first-principles investigation of the physical properties of $Bi_2Te_2Se$ topological insulator (2020). arXiv:2005.13393. |
| [68] | M.E. Fine, L.D. Brown, H.L. Marcus, Elastic constants versus melting temperature in metals, Scr. Metall. 18 (1984) 951-956. https://doi.org/10.1016/0036-9748(84)90267-9. |
| [69] | W.L. McMillan, Transition temperature of strong-coupled superconductors, Phys. Rev. 167 (1968) 331. https://doi.org/10.1103/PhysRev.167.331. |
| [70] | J.J. Hopfield, Angular momentum and transition-metal superconductivity, Phys. Rev. 186 (1969) 443. https://doi.org/10.1103/PhysRev.186.443. |
| [71] | J.P. Carbotte, Properties of boson-exchange superconductors, Rev. Mod. Phys. 62 (1990) 1027. https://doi.org/10.1103/RevModPhys.62.1027. |
| [72] | A.K.M.A. Islam, S.H. Naqib, Possible explanation of high-$T_c$ in some 2D cuprate superconductors, J. Phys. Chem. Solids 58 (1997) 1153. https://doi.org/10.1016/S0022-3697(96)00145-X. |
| [73] | M.A. Blanco, E. Francisco, V. Luana, GIBBS: isothermal-isobaric thermodynamics of solids from energy curves using a quasi-harmonic Debye model, Comput. Phys. Commun. 158 (2004) 57–72. doi:http://dx.doi.org/10.1016/j.comphy.2003.12.001. |

**Author Contributions**

F.P. performed the theoretical calculations and contributed in the analysis. S.H.N. designed the project, analyzed the results, and wrote the manuscript. Both the authors reviewed the manuscript.

**Additional Information**

**Competing Interests**

The authors declare no competing interests.